\documentclass[a4paper,12pt]{article}
\usepackage{epsfig}
\usepackage{psfig}
\usepackage{citesort}
\date{\today}
\def\unit{\leavevmode\hbox{\small1\kern-3.6pt\normalsize1}}
\newcommand{\be}{\begin{equation}}
\newcommand{\ee}{\end{equation}}
\newcommand{\bea}{\begin{eqnarray}}
\newcommand{\eea}{\end{eqnarray}}
\def\ie{{\it i.e.}}
\def \ben{\begin{enumerate}}
\def \een{\end{enumerate}}
\def \bit{\begin{itemize}}
\def \eit{\end{itemize}}

\def\lsim{\raise0.3ex\hbox{$\;<$\kern-0.75em\raise-1.1ex\hbox{$\sim\;$}}}
\def\gsim{\raise0.3ex\hbox{$\;>$\kern-0.75em\raise-1.1ex\hbox{$\sim\;$}}}
\def\Frac#1#2{\frac{\displaystyle{#1}}{\displaystyle{#2}}}
\def \av#1{\left\langle #1\right\rangle}

\begin{document}
\pagestyle{empty}  
\begin{flushright}
\small
IPPP/02/14\\
DCPT/02/28\\
\end{flushright}
\renewcommand{\thefootnote}{\fnsymbol{footnote}}
\renewcommand{\baselinestretch}{1.2} \large\normalsize
\vspace{.3cm}
\begin{center}
{\bf{\Large  
CP violation in supersymmetric models with Hermitian Yukawa couplings
and $A$-terms}}\\
\vspace* {1cm}
{\large\normalsize Shaaban Khalil}\\
\vspace* {5mm}
{\it IPPP, Physics Department, Durham University, DH1 3LE,
Durham, U.K}\\
\vspace* {2mm}
{\it Ain Shams University, Faculty of Science, 
Cairo, 11566, Egypt}
\end{center}   

\vspace*{10mm}
\begin{minipage}[h]{14.0cm}
\begin{center}
\large{\bf Abstract}\\[3mm]
\end{center} 
We analyse the CP violation in supersymmetric models
with Hermitian Yukawa and trilinear couplings. We show that Hermitian 
Yukawa matrices can be implemented in supersymmetric models with
$SU(3)$ flavor symmetry. An important feature of this class of models 
is that the supersymmetric contributions to the EDM of the neutron 
and mercury atom are suppressed. 
In this framework, $\varepsilon_K$ can be saturated by a small 
non--universality of the soft scalar masses through the gluino contribution. 
We perform a detailed analysis for the supersymmetric contributions to 
$\varepsilon'/\varepsilon$. Although, the gluino contribution is found to 
be negligible due to a severe cancellation between LR and RL mass insertions, 
the chargino contribution can be significant and accommodate 
the experimental results. Additionally, we find that the SUSY contributions 
to $\varepsilon'/\varepsilon$ from the effective $\bar{s}dZ$ vertex and
the $\Delta I =3/2$ operators are insignificant. 
We point out that the standard model gives the 
leading contribution to the CP asymmetry in $B \to \psi K_S$ decay, while
the dominant chargino contribution to this asymmetry is $\lsim 0.2$. Thus,
no constraint is set on the non--universality of this model by the recent
BaBar and Belle measurements.

\end{minipage}   
\newpage
\setcounter{page}{2}
\pagestyle{plain}
\renewcommand{\thefootnote}{\arabic{footnote}}
\setcounter{footnote}{0}
%
\section{\bf \large  Introduction}
CP violation is one of the outstanding problems in high energy physics.
Although the standard model (SM) is able (till now) to accommodate
the experimentally observed CP violation, there are strong hints
of additional sources of CP violation 
beyond the phase in the Cabibbo-Kobayashi-Maskawa matrix
($\delta_{\mathrm{CKM}}$). The strongest motivation for this suggestion
is that the strength of CP violation in SM is not sufficient to explain 
the cosmological baryon asymmetry of our universe~\cite{BAU}.

In supersymmetric (SUSY) extensions of the SM, there are new CP violating 
phases which arise from the complexity of the soft SUSY breaking terms and
the SUSY preserving $\mu$-parameter. These new phases have significant 
implications and can modify the SM predictions in CP violating phenomena. 
In particular, they would give large contributions to the electric dipole 
moment (EDM) of the neutron and mercury atom~\cite{abel-edm}, 
to CP violating 
parameters ($\varepsilon$ and $\varepsilon'$) of $K-\bar{K}$ 
system~\cite{frere,masiero,nonuniv}, and to the CP asymmetries 
in the $B-\bar{B}$ system~\cite{Goto,neubert2,Bailin,Ali,masiero2,buras4}. 
These phases can be 
classified into two categories. 
The first category includes  flavor-independent phases such as the phases 
of the $\mu$-parameter, $B$-parameter, gaugino masses and the overall 
phase of the trilinear couplings. The other category includes 
the flavor-dependent phases, i.e. the phases of the off--diagonal elements 
of $A_{ij}$ and phases in the squark mass matrix $ m_{ij}^2$.  
Two of the flavor-independent phases can be eliminated by the $U(1)_R$ 
and $U(1)_{PQ}$ transformations. 

However, the non--observation of EDMs imposes a stringent constraint on 
flavor--independent SUSY phases, the so--called SUSY CP problem. 
Four solutions to this problem have been indicated so far. In the first, the 
CP is an approximate symmetry and therefore the CP violating phases 
are very small ($\sim 10^{-2}$). However, the large CP asymmetry in 
$B \to \psi K_s$ decay observed by BaBar~\cite{2beta:babar} and 
Belle~\cite{2beta:belle} implies that CP is significantly violated and 
disfavors this possibility. In the second, the sfermion of the first two 
generations are very heavy while the third generation remains light. 
However, it was shown
~\cite{abel-edm} that in order to satisfy the EDM of the mercury atom the
sfermion masses have to be of order 10 TeV, which leads to a large hierarchy 
between SUSY and electroweak scales. A third possibility is that the full
computation of SUSY contributions to the EDMs involves accidental 
cancellations among various contributions which may allow for regions
of parameter space with large phases. However, it has been found 
~\cite{abel-edm} that such EDM cancellation can not occur simultaneously for 
the electron, neutron and mercury. Finally, 
the SUSY CP violation can have a flavor off-diagonal character just as in the 
SM~\cite{ABKL}. In this latter possibility, the origin of the CP violation is 
assumed to be closely related to the origin of flavor structures 
rather than the origin of SUSY breaking. Thus
the flavor blind quantities as the $\mu$--term and gaugino masses
are real. 

The class of models with flavor--dependent CP violation are based on two 
major assumptions. First, the SUSY breaking does not break CP, 
which can happen in some explicit string models~\cite{morris}. 
Second, the flavor structure of the model is 
Hermitian, \ie, the Yukawa matrices and the $A$--terms have to be 
Hermitian, in order to ensure that the diagonal elements of the $A$--terms
are real in any basis and do not induce unacceptably large EDMs. 

In Ref.~\cite{ABKL}, it was shown that within these assumptions, the 
supersymmetric CP problem can naturally be resolved and a correlation
between the CP asymmetry of the $B \to X_s \gamma$ decay and the EDMs
is predicted. However, it was also found that SUSY contribution to 
$\varepsilon_K$ is, in general, very small and also that the dominant 
gluino contribution to $\varepsilon'/\varepsilon$ is negligible 
due to a cancellation between the contributions involving 
$(\delta_{12}^d)_{LR}$ and  $(\delta_{12}^d)_{RL}$ mass insertions. 
In fact, the SM prediction for $\varepsilon_K$ can be fitted with the 
current experimental data, however, due to the large uncertainties  
in the theoretical estimate of 
$\varepsilon'/\varepsilon$ it is rather unclear if the SM result is 
consistent with the observed measurements by KTeV~\cite{KTeV} and 
NA31~\cite{NA31}. Thus, it is necessary systematically to 
analyse the different SUSY contributions to
$\varepsilon'/\varepsilon$ to show if it is possible to saturate 
the observed value in SUSY models with Hermitian Yukawa couplings and 
$A$--terms.

In this paper, we study more explicitly CP violation in the K and 
B system due to flavor dependent phases in SUSY models with the 
Hermiticity assumption. We show that, in order to saturate 
$\varepsilon_K$ in a viable region of parameter space non--universality 
between the squark soft masses is required. This non--universality is also 
essential to enhance the chargino contribution to $\varepsilon'/\varepsilon$. 
We demonstrate that the SUSY flavor off--diagonal phases have significant 
implications on the direct CP asymmetry in the $B\to X_S
\gamma$ decay, while their effect on the CP asymmetry in the $B\to \psi K_S$ 
decay are negligible.

The paper is organized in the following way. In section 2 we start by 
emphasizing the possibility of obtaining Hermitian Yukawa couplings in 
SUSY models with $SU(3)$ flavor symmetry, and then show that the EDMs 
in this class of models are one or two order of magnitude below 
the experimental constraints without any fine--tunning.
Section 3 is devoted to the study of CP violation in the Kaon system.
In section 4 we consider the CP violation in the $B$--sector and 
show that in this framework the large CP asymmetry in the $B\to \psi K_S$ 
decay is given by the SM contribution while the SUSY contribution is 
very small. In contrast, the SUSY contribution to the CP asymmetry in 
the $B\to X_S \gamma$ decay can be as large as $\pm 10\%$.
Finally, the conclusions are presented in section 5. 

\newpage
%
\section{\bf \large Hermiticity and EDM suppression}
As discussed in the introduction, an elegant solution for 
suppressing the EDMs in SUSY models is to have Hermitian flavor 
structures, \ie\ $Y^a = Y^{a^{\dag}}$, and 
$A^a = A^{a^{\dag}}$. It is known that Hermitian Yukawa matrices 
can be implemented in models with left--right symmetry~\cite{mohapatra} and 
horizontal flavor symmetry~\cite{yanagida}. In Ref.\cite{ABKL}, 
it was assumed that Hermitian Yukawa appeared due to a horizontal 
symmetry $U(3)_H$ which gets broken 
spontaneously by the VEVs of the real adjoint fields, $T^a$, hence
the Yukawa couplings are given as $Y_{ij} = \frac{g_H}{M}
\av{T^a} (\lambda^a)_{ij}$, where $g_H$ is a coupling constant 
of order one, $M$ is a mass scale much higher than the electroweak
scale, $\lambda^a$, for $a=1,..,8$ are Gell-Mann matrices
and $\lambda^0$ is proportional to the unit matrix. However, the real
fields $T^a$ may only arise from non--supersymmetric sector.  
In fact, if $T^a$ are the scalar components of chiral multiplets, 
they are intrinsically complex and Hermitian Yukawas arise only if the 
VEVs of $T^a$ are real.

As we will show in the following subsection, it is possible to obtain 
Hermitian Yukawa couplings in SUSY models with $SU(3)$ flavor symmetry
broken by complex VEVs of scalar fields $\phi_a$ and 
$\bar{\phi}^a$ in the triplet and antitriplet representation 
of $SU(3)$ respectively.
The local $SU(3)$ flavor symmetry provides a dynamical origin for the observed 
fermion masses and a natural explanation for three quark-lepton
families. A considerable amount of work has been done concerning the 
implications of this symmetry on solving the 
fermion flavor problem~\cite{Ross}.
  
%
\subsection{\bf \normalsize Hermitian Yukawa from $SU(3)$ flavor symmetry}
Here, we show that Hermitian Yukawa couplings can be motivated by
supersymmetric models with flavor symmetry $SU(3)_F$. We consider a
SUSY model with the gauge group $G_{SM} \times SU(3)_F$, where
$G_{SM}$ refers to the standard model gauge group.
Under $SU(3)_F$, the matter content of the MSSM 
is assigned the following quantum numbers:
\begin{equation}
\{ Q_a , L_a \} \equiv 3 \quad \mathrm{and} \quad
\{ u^c_a ,\; d^c_a,\; e^c_a\} \equiv \bar{3}\;,
\label{A1}
\end{equation}
while the MSSM Higgs are singlets under the $SU(3)_F$ and 
have the charges $\{ H_u , H_d \} \equiv 1$. 
The extra Higgs fields that are used to break $SU(3)_F$ are
$\phi_a \equiv 3 $ and 
$\bar{\phi}^a \equiv \bar{3}$, $a=1,2,3$.

In this class of models, the lowest dimensional $SU(3)_{F}$ invariant
operators in the superpotential, which are 
responsible for generating the fermion masses, are given by
\begin{equation}
W_{\mathrm{Yuk}} = h_u Q_a u^c_b H_u \frac{\bar{\phi}^a \phi_b}{M^2} + h_d Q_a
d^c_b H_d \frac{\bar{\phi}^a \phi_b}{M^2}+ h_l L_a e^c_b H_d
\frac{\bar{\phi}^a \phi_b}{M^2}.
\label{A2}
\end{equation}
Thus for $\av{\bar{\phi}}=\av{\phi}^*$, \ie, 
$\av{\phi_a}= v_a e^{i\varphi_a}$ and  $\av{\bar{\phi}_b}= 
v_b e^{-i\varphi_b}$, the Yukawa couplings are given by
\begin{equation}
Y_{ab}^u = h_u \frac{v_a v_b}{M^2} \;e^{i (\varphi_a - \varphi_b)}\;. 
\label{A3}
\end{equation}
Similar expressions hold for $Y^d$
and $Y^l$. Eq.(\ref{A3}) clearly displays the usual form
for Hermitian Yukawa couplings. Now let us discuss briefly the 
minimization of the scalar potential of the $\phi$ fields. 
The most general renormalizable superpotential involving these
Higgs fields has the form
\begin{equation}
W = \mu \phi_a \bar{\phi}^a + \lambda \phi_a \bar{\phi}^a S + W'(S)\;,
\end{equation}
where the $S$ field is a singlet under both $G_{SM}$ and $SU(3)_F$.
The requirement that the $\phi$'s and $S$ fields do not contribute to SUSY 
breaking implies that $F_{\phi} = F_S = 0$. The scalar potential is given by
\begin{equation}
V = \left\vert \mu \bar{\phi}^a +\lambda \bar{\phi}^a S \right\vert^2
+ \left\vert \mu \phi^a + \lambda \phi^a S \right\vert^2 + \left\vert
\lambda \phi_a \bar{\phi}^a + \frac{\partial W'} {\partial S}
\right\vert^2 + V_D + V_{SB}\;,
\end{equation}
where 
\begin{equation} 
V_D = \frac{g^2_{3H}}{2}\sum_b \left(\phi^\dagger_a T^b \phi_a + 
\bar{\phi}^\dagger_a T^b \bar{\phi}_a \right)^2\;. 
\end{equation} 
Note that $\phi^\dagger_a$ is an anti-triplet under $SU(3)_F$ 
as well as $\bar{\phi}_a$ so the above potential is $SU(3)_F$ invariant.
In the above equation, $T^b$ are the generators of the $SU(3)_F$ group,
and the sum extends over all these generators.  
Finally we assume the following soft SUSY breaking terms
\begin{equation} 
V_{SB} = m_{\phi_a}^2 \vert \phi_a
\vert^2 +m_{\bar{\phi}_a}^2 \vert \bar{\phi}_a \vert^2 + \left[ A_{\phi_a}
\phi_a \bar{\phi}^a S + B_{\phi_a} \phi_a \bar{\phi}^a +\mathrm{h.c.}\right]\;.  
\end{equation}
The minimization of the scalar potential with respect to $\phi_a$ and
$\bar{\phi}^a$ depends on the soft SUSY breaking terms and, for 
a particular choice of these parameters, one can obtain the following 
VEV's for $\phi_a$ and $\bar{\phi}_b$
\begin{equation}
\av{\phi_a}= v_a  e^{i \varphi_a}\; ,\quad  
\av{\bar{\phi}_b}= v_b e^{-i\varphi_b}\;, 
\end{equation}
as required in order to get Hermitian Yukawa textures. 

Furthermore, since $\bar{\phi}^a \phi_a$ is a singlet under both 
the $G_{SM}$ and $SU(3)_F$, it can couple to $H_u$ and $H_d$ to generate
the $\mu$--term. In this case we can have the following leading term
in the superpotential:
\begin{equation}
W_{\mu} \sim \frac{\bar{\phi}^a \phi_a}{M} H_u H_d. 
\end{equation}
Thus, the $\mu$--term will be given, after the $SU(3)_F$ is 
spontaneously broken,
by $\mu = v_a^2/M$ which is real. However the $SU(3)_F$ symmetry, like 
any other flavor symmetry and left--right symmetry, can not guarantee 
the reality of the gaugino masses and we have to make the additional assumption
that the SUSY breaking dynamics conserves CP. This seems natural if CP breaking
is associated with the origin of the flavor structure~\cite{ABKL}.  
%
\subsection{\bf \normalsize EDM--free SUSY CP violating phases}
We have shown that Hermitian Yukawa matrices and a real $\mu$--term
can arise naturally in models with $SU(3)_F$ symmetry and the assumption 
that CP violation and SUSY breaking have different origins leads to 
$\mathrm{arg}(M_i)=0$ where $M_i$ are the gaugino masses.
In this case the $A$--terms can also be Hermitian and the EDM 
problem is naturally avoided~\cite{ABKL,mohapatra}. 

In supergravity models, the trilinear parameters are given in 
terms of the K\"ahler potential and the Yukawa couplings 
\begin{eqnarray}
A_{\alpha \beta \gamma} = F^m \bigl[ \hat K_m  + \partial_m \ln
Y_{\alpha \beta \gamma} -\partial_m \ln (\tilde K_{\alpha} \tilde K_{\beta} 
\tilde K_{\gamma} ) \bigr] \;,
\label{A-terms}
\end{eqnarray}
where the Latin indices refer to the hidden sector fields that break 
SUSY and the Greek indices refer to the observable fields. 
According to our previous assumption, the $F^m$ is real. Also $\tilde 
K_{\alpha}$ and $\hat K_m$ are always real, thus the $A$--terms are 
Hermitian if the derivatives of the K\"ahler potential are either 
generation--independent or the same for the left and right fields of the 
same generation, \ie, if $\tilde{K}_{Q_{L_i}} \tilde{K}_{U_{R_j}}=
\tilde{K}_{Q_{L_j}} \tilde{K}_{U_{R_i}}$. 
These conditions are usually satisfied in string models.
It is interesting to note that although the SUSY breaking does not bring in
new source of CP violating, the trilinear soft parameters involve 
off--diagonal CP violating phases of $\mathcal{O}(1)$. This stems from the 
contribution of the term $\partial_m \ln Y_{\alpha\beta\gamma}$, which
has been found to be significant and sometimes even dominant in string 
models~\cite{stringcp}.
In what follows, we will show that these phases are unconstrained by the 
EDMs and will study their phenomenological implications in the $K$ and $B$ 
systems.

The relevant quantities appearing in the soft Lagrangian are $(Y^A_q)_{ij}=
(Y_q)_{ij} (A_q)_{ij}$ (indices not summed) which are also Hermitian at 
the GUT scale. For the sake of definiteness, we consider the following 
Hermitian Yukawa matrices at the GUT scale
\begin{eqnarray}\label{yukawa}
Y^u &= & \lambda_u \left(
\begin{array}{ccc}
5.94\times10^{-4} & 10^{-3}~i & -2.03\times10^{-2}  \\
- 10^{-3}~i & 5.07\times10^{-3} & 2.03\times10^{-5}~i \\
-2.03\times10^{-2} & -2.03\times10^{-5}~i & 1 
\end{array} \right) \;,\\
Y^d &=& \lambda_d \left(
\begin{array}{ccc}
6.84 \times 10^{-3} & (1.05+0.947~i) \times 10^{-2} & -0.023 \\
(1.05-0.947~i)\times 10^{-2} & 0.0489 & 0.0368~i\\
-0.023 & -0.0368~i & 1 
\end{array}\right),
\end{eqnarray}
where $\lambda_u=m_t/v \sin\beta$ and $\lambda_d=m_b/v\cos\beta$. 
These matrices reproduce, at low energy, the quark masses and the CKM
matrix. The renormalization group (RG) evolution of Yukawa couplings 
and the $A$ terms slightly violate the Hermiticity. Therefore, the resulting 
$Y^A_q$ at the electroweak scale has very small non--zero phases
in the diagonal elements (due to the large suppression from the 
off--diagonal entries of the Yukawa). 
However, what matters is the relevant phases appearing in the squark mass 
insertions in the super-CKM basis, \ie, the basis where the Yukawa matrices 
are diagonalized by a unitary transformation of the quark superfields 
$\hat{U}_{L,R}$ and $\hat{D}_{L,R}$ 
(Note that since the Yukawas are Hermitian matrices they are diagonalized
by one unitary transformation,\ie $V^q_L=V^q_R$):
\bea
\hat{U}_{L,R}&\to&V^u~\hat{U}_{L,R},~ \hat{D}_{L,R}\to V^d~\hat{D}_{L,R},
\nonumber\\
Y^u &\to& V^{u^T}~ Y^u~ V^{u^*}\equiv \mathrm{diag}(h_u,h_c,h_t),\nonumber\\ 
Y^d &\to& V^{d^T}~ Y^d~ V^{d^*}\equiv \mathrm{diag}(h_d,h_s,h_b).
\eea
Accordingly the trilinear terms $Y^A_q$ transform as $Y^A_q \to
V^{q^T} Y^A_q V^{q^*}$. Thus the $Y^A_q$ stay Hermitian to a very good degree
in the super-CKM basis. Therefore, the imaginary parts of the flavor conserving
mass insertions 
\begin{equation}
(\delta^{d(u)}_{ii})_{LR} = \frac{1}{m_{\tilde{q}}^2}
\left[ \left(V^{q^T} Y^A_q V^{q^*}\right)_{ii} v_{1(2)} -
\mu Y^{d(u)}_i v_2(1) \right],
\end{equation}
that appear in the EDM calculations are suppressed. In the above
formula the $m_{\tilde{q}}$ refers to the average squark mass and 
$v_i=\av{H_i^0}/\sqrt{2}$.

The effective Hamiltonian for the EDM of a fermion $f$ containing
dimension-5 and 6 operators is given by
~\cite{kane-edm}
\begin{equation}
H^{\mathrm{EDM}}_{\mathrm{eff}} = \sum_i C_i (\mu) \mathcal{O}_i + h.c.\;,
\end{equation}
where $\mathcal{O}_i$ are given by
\bea
\mathcal{O}_1=-\frac{i}{2} \bar{f} \sigma_{\mu\nu} \gamma_5 f F_{\mu\nu},~~~~~
\mathcal{O}_2=-\frac{i}{2} \bar{f} \sigma_{\mu\nu} \gamma_5 f 
G^a_{\mu\nu},~~~~~ 
\mathcal{O}_3=-\frac{1}{6} f_{abc} G^a_{\mu\rho} G_{\nu}^{b\rho}
G^c_{\lambda\sigma} \epsilon^{\mu\nu\lambda\sigma}.
\eea
$\mathcal{O}_{1,2}$ refer to the electric and chromoelectric dipole moment
operators and $\mathcal{O}_3$ to the Weinberg three gluino operator.
All these operators can contribute to the quark EDM while only 
$\mathcal{O}_1$ contributes to the electron EDM, \ie, 
the Wilson coefficients $C^e_2$ and $C^e_3$ of the electron are identically
zero. The supersymmetric contributions to the Wilson coefficients of the
quark result from the 1-loop gluino, chargino, and neutralino exchange
diagrams and also the 2-loop gluino--quark--squark diagram. As emphasized
in Ref.\cite{abel-edm}, the most stringent constraint on the SUSY CP violating 
phases comes from the recent experimental bounds on the EDMs of the 
neutron and mercury atom. Therefore we will not discuss the electron EDM here.

The EDMs of quarks, using the naive dimension analysis, are given by
\begin{equation}
d_q = \eta_1~ C_1 + \frac{e}{4\pi}~ \eta_2~ C_2 + \frac{e \Lambda}{4\pi}
\eta_3~ C_3,
\end{equation} 
where the QCD correction factors are $\eta_1=1.53$,
$\eta_2 \simeq \eta_3\simeq 3.4$, and $\Lambda\simeq 1.19$ GeV is 
the chiral symmetry breaking scale. The dominant 1-loop gluino 
contributions to the Wilson coefficients of the down and up
quarks are given by
\bea
C^{d(u)}_1 &=&-\frac{2}{3}~ \frac{e \alpha_s}{\pi}~ Q_{d(u)}~ 
\frac{m_{\tilde{g}}} {m_{\tilde{q}}^2}~ \mathrm{Im}(\delta^{d(u)}_{11})_{LR}
~ M_1(x),\\
C^{d(u)}_2 &=&-\frac{1}{4}~ \frac{g_s \alpha_s}{\pi}~ \frac{m_{\tilde{g}}}
{m_{\tilde{q}}^2}~ \mathrm{Im}(\delta^{d(u)}_{11})_{LR}~ M_2(x).
\eea
Here $m_{\tilde{g}}$ is the gluino mass and the function $M_1(x)$ is 
defined by
\bea
M_1(x) &=& \frac{1+4 x - 5 x^2 + 4 x \ln(x) + 2 x^2
\ln(x)}{2(1-x)^4}\;,\\
M_2(x) &=& \frac{22-20 x - 2 x^2 + 9 \ln(x) + 16 x \ln(x) - x^2 \ln(x)}
{3(1-x)^4}\;,
\eea
with $x= m_{\tilde{g}}^2/m_{\tilde{q}}^2$. In the quark 
model, the EDM of the neutron is given by $d_n= \frac{1}{3} (4 d_d -d_u)$
and the current experimental bound~\cite{nedm}
\begin{equation}
d_n < 6.3 \times 10^{-26}\; \mathrm{e\ cm} \quad (90\% \mathrm{C.L.})
\end{equation}
leads to the constraint $\mathrm{Im}(\delta^{d(u)}_{11})_{LR} 
\lsim 10^{-6} - 10^{-7}$. However,
it turns out that the recent experimental limit on the EDM of the 
mercury atom~\cite{mercury}
\begin{equation}
d_{H_g} < 2.1 \times 10^{-28}\; \mathrm{e\ cm},
\label{edm:mercury}
\end{equation}
implies stronger bounds on these mass insertions (more than 
an order of magnitude more stringent than those imposed by the EDM of the 
neutron) and in addition to 
$\mathrm{Im}(\delta^{d}_{22})_{LR} \lsim 10^{-5} - 10^{-6}$
~\cite{abel-edm}, due considerable contributions from the strange quark
to the mercury EDM. Recall that the mercury EDM is sensitive to 
the chromoelectric EDM of quarks ($C_2^q$) and the limit in 
Eq.(\ref{edm:mercury}) can be translated into 
$\vert C^d_2 - C_2^u - 0.012 C_2^s \vert /g_s < 7 \times 10^{-27} cm$
~\cite{olive}.

We start our analysis by revisiting the EDM constraints on the flavor
off diagonal phases of SUSY models with Hermitian Yukawa as in 
Eq.(\ref{yukawa}) and the following Hermitian $A$--terms:
\begin{equation}
A_d=A_u = \left( \begin{array}{ccc}
A_{11} & A_{12}~e^{i\varphi_{12}} & A_{13}~e^{i\varphi_{13}}  \\
A_{12}~e^{-i\varphi_{12}} & A_{22} &A_{23}~e^{i\varphi_{23}}  \\
A_{13}~e^{-i\varphi_{13}}&A_{23}~e^{-i\varphi_{23}}  & A_{33} 
\end{array} \right) \;.
\label{soft1}
\end{equation}
We also assume that the soft scalar masses and gaugino masses $M_a$
are given by
\bea
M_a &=& m_{1/2} ,~~ a=1,2,3,\\
m^2_Q &=&m_{H_1}^2 = m_{H_2}^2=m_0^2,\\
m^2_U &=& m^2_D=m_0^2~ \mathrm{diag}\{1,\delta_1,\delta_2\},
\label{soft2}
\eea
where the parameters $\delta_i$ and $A_{ij}$ can vary in the ranges
$[0,1]$ and $[-3,3]$ respectively. Note that in most string inspired 
models, the squark mass matrices are diagonal but not necessary
universal. The non--universality of the squark masses is not constrained
by the EDMs. However, this non--universality (specially between the 
first two generations of the squark doublets) is  severely constrained by 
$\Delta M_K$ and $\varepsilon_K$.

\begin{figure}[t]
\begin{center}
\hspace*{-7mm}
\epsfig{file=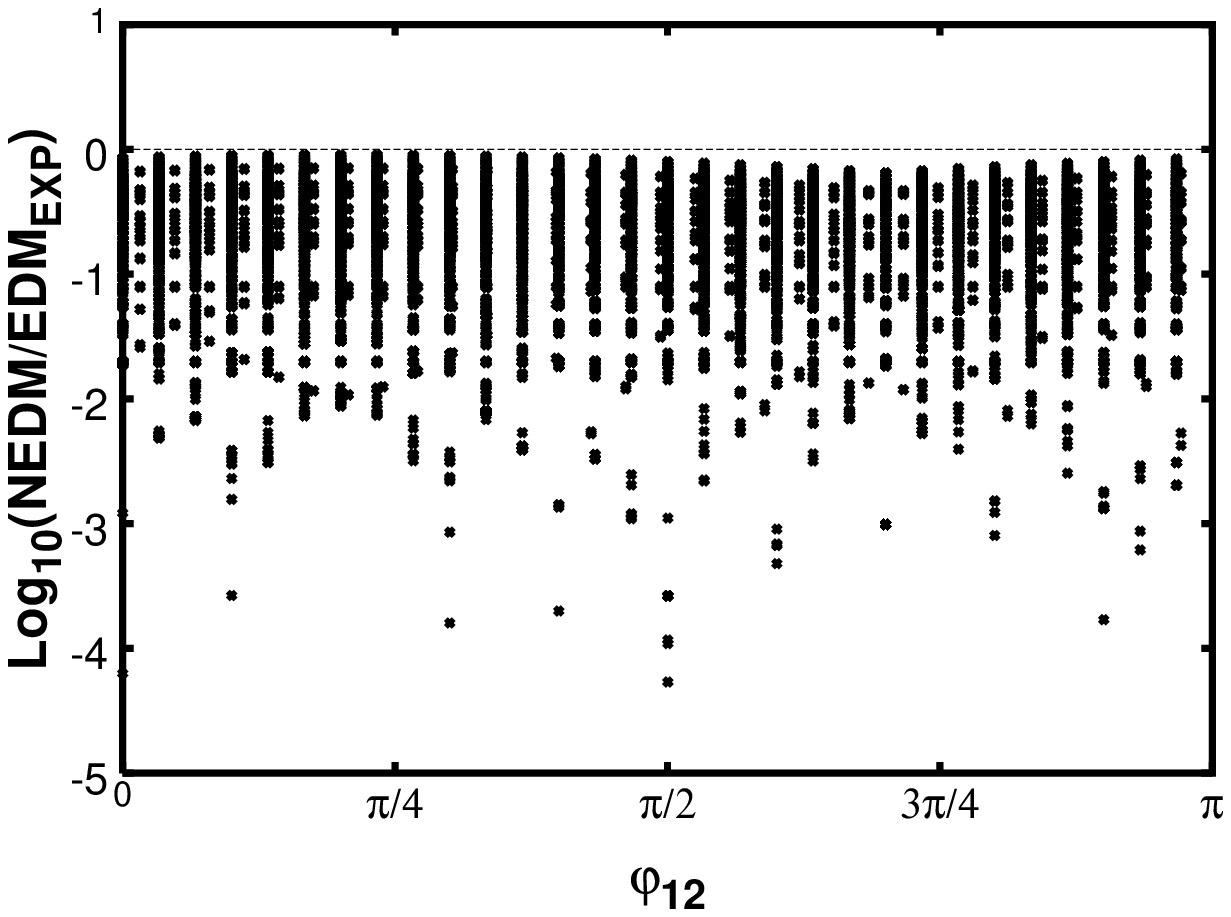,width=8cm,height=6.25cm} \quad
\epsfig{file=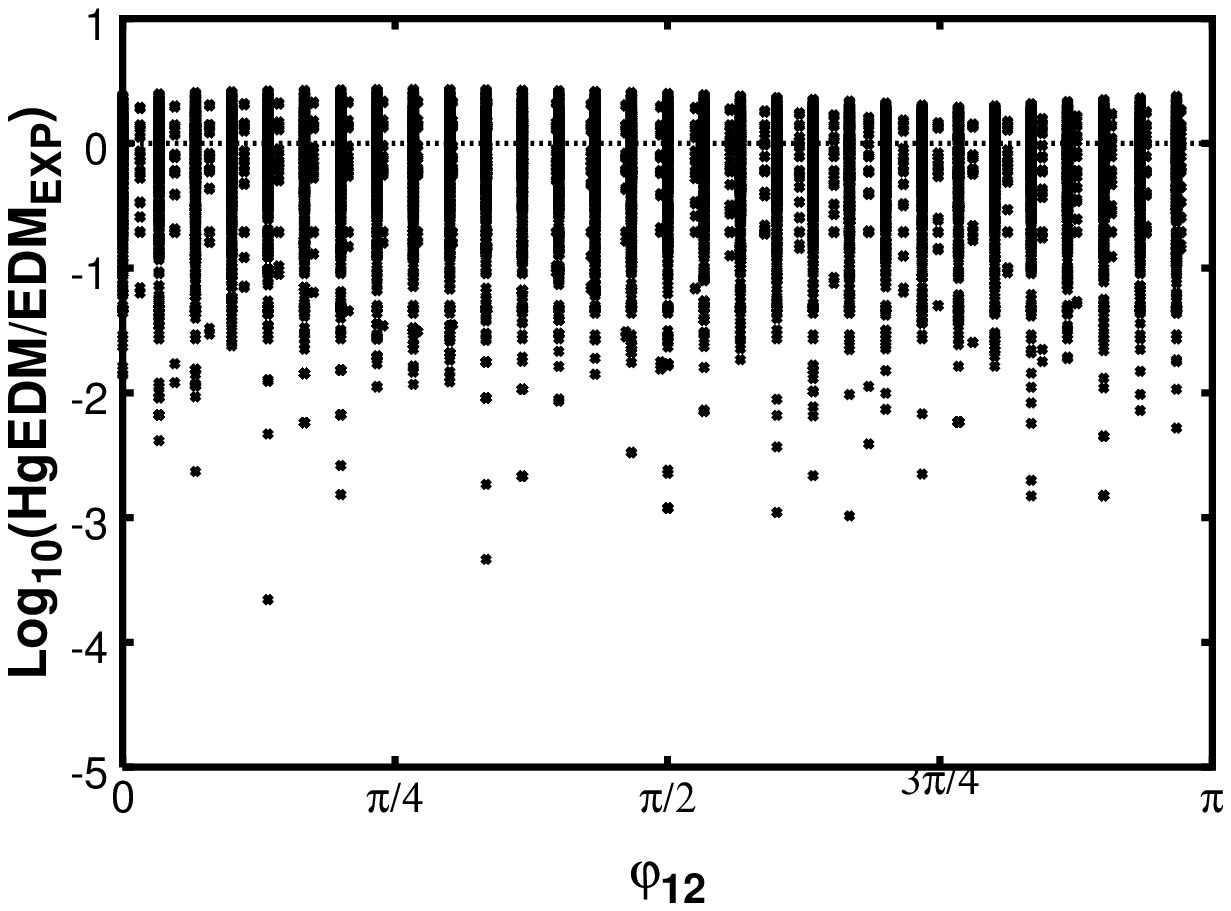,width=8cm,height=6.25cm}\\
\caption{{\small Neutron and mercury EDMs versus $\varphi_{12}$ 
for  $\tan\beta=5$ and $m_0=m_{1/2}=200$ GeV.}}
\label{fig1}
\end{center}
\end{figure}

In Fig. 1 we display scatter plots for the neutron and mercury EDMs 
versus the phase $\varphi_{12}$ for $\tan \beta=5$, $m_0=m_{1/2}=200$ GeV,
$A_{ij}$ are scanned in the range $[-3,3]$, 
and the phases $\varphi_{13}$ and $\varphi_{23}$ are
randomly selected in the range $[0,\pi]$. As stated above, the parameters 
$\delta_i$ are irrelevant for the EDM calculations and we set them here 
to one. Finally, since $\mu$ is real the EDM results
display very little dependence on $\tan \beta$. 

It is important to mention that we have also imposed the constraints 
which come from the requirement of absence of charge and colour breaking 
minima as well as the requirement that the scalar potential be bounded
from below~\cite{CCB}. These conditions may be automatically
satisfied in minimal SUSY models, however in models with non--universal
$A$--terms they have to be explicitly checked. In fact, sometimes these
constraints are even stronger than the usual bounds set by the flavor
changing neutral currents~\cite{casas}.

As can be seen from Fig. 1, the EDMs do not exceed the experimental bounds 
for most of the parameter space. Generally, they are one or two order of 
magnitude below the present limit, and the flavor--off diagonal phases of the 
$A$--terms can be $\mathcal{O}(1)$ without fine--tunning.
The points that lead to mercury EDM above the experimental bound 
correspond to $\varphi_{23}\simeq \pi/2$. This phase induces a considerable 
contribution to the chromoelectric EDM of the strange quark $C_2^s$. 
Thus the compatibility with mercury EDM experiment requires that the phase 
$\varphi_{23}$ should be slightly smaller than $\pi/2$. 
 

%
\section{\bf \large CP violation in the Kaon system}
We have shown in the previous section that the Hermiticity of 
the Yukawa couplings and $A$--terms allows the existence of large 
off--diagonal SUSY CP violating phases while keeping the EDMs sufficiently 
small. However, the important question to address is whether these
``EDM--free'' phases can have any implication
on other CP violation experiments. In this section, we will 
concentrate on possible effects in the kaon system. 

Recently, it has been pointed out that, in SUSY models with generic 
non--degenerate $A$--terms (where the phases of the diagonal
elements are set to be very small by hand in any basis to 
satisfy the EDM bounds), it is possible to have large effects
in CP violation observables, in particular $\varepsilon_K$ and 
$\varepsilon'/\varepsilon$~\cite{frere,masiero,nonuniv}. However, as we will
show below, in the Hermitian scenario the situation is quite different 
and it is not straightforward 
to realize the above mentioned mechanism (which relies on the LR
down squark mass insertions) to obtain significant SUSY contribution to 
$\varepsilon_K$ and $\varepsilon'/\varepsilon$. As shown in Ref.\cite{ABKL},
the typical values of $\varepsilon_K$ in this class of models are smaller 
than the experimental measurement (even with very large off--diagonal elements,
$A_{ij}\sim 5 m_0$). Moreover, it turns out that 
$\varepsilon'/\varepsilon$ is also very small ($\sim 10^{-6}$) due to a 
severe cancellation between the different contributions. In the following, 
we will show that we have to consider the other SUSY contributions from the LL 
and RR sectors in order to saturate both $\varepsilon_K$ and 
$\varepsilon'/\varepsilon$.
%
\subsection{\bf \normalsize Indirect CP violation}

In the kaon system and due to a CP violation in $K^0-\bar K^0$ mixing, the 
neutral kaon mass eigenstates are superpositions of CP--even ($K_S$) and
CP--odd ($K_L$) components. However, the CP--odd $K_L$ decays into two pions 
through its small CP--even component. This decay, $K_L \to \pi \pi$, was the 
the first observation of CP violation. The measure for the indirect CP
violation is defined as 
\be
\varepsilon_K = {A(K_L\rightarrow \pi\pi)\over A(K_S\rightarrow
\pi\pi) }\;.
\ee
The experimental value for this parameter is 
$\varepsilon_K \simeq 2.28 \times 10^{-3}$.  Generally, 
$\varepsilon_K$ can be  calculated via
\be
\varepsilon_K = {1\over \sqrt{2} \Delta M_K} {\rm Im} \langle K^0
\vert H^{\Delta S=2}_{\rm eff} \vert \bar K^0 \rangle \;.
\label{deltamk}
\ee
Here $H^{\Delta S=2}_{\rm eff}$ is the effective Hamiltonian for
the $\Delta S=2$ transition. It can be expressed via the Operator
Product Expansion as
\begin{equation}
H^{\Delta S=2}_{\rm eff}=\sum_{i} C_i(\mu) Q_i + h.c. \;,
\end{equation}
where $C_i(\mu)$ are the Wilson coefficients and $Q_i$ are the
relevant local operators, which are given in Ref.\cite{ciuchini}.
The main uncertainty in this calculation arises from
the matrix elements of $Q_i$, whereas the Wilson coefficients can
be reliably calculated at high energies and evolved down to low
energies via the RG running.

The $K^0-\bar{K}^0$ transition can be generated through the box diagrams
with  W, Higgs, neutralino, gluino, and chargino exchange. 
The off--diagonal entry in the kaon 
mass matrix,  $\mathcal{M}_{12}=\langle K^0 \vert H^{\Delta S=2}_{\rm
eff} \vert \bar K^0 \rangle$, is given by
\be
\mathcal{M}_{12} = \mathcal{M}_{12}^{\mathrm{SM}} +  
\mathcal{M}_{12}^{\mathrm{H}}+ \mathcal{M}_{12}^{{\chi^0 }}+
\mathcal{M}_{12}^{{\tilde g}} +  \mathcal{M}_{12}^{{\chi^+}}.
\ee
The SM contribution can be written as~\cite{buras1}
\begin{equation}
\mathcal{M}^{\mathrm{SM}}_{12} = \frac{G_F^2 M^2_W}{12 \pi^2} 
F^2_K M_K \hat{B}_K \mathcal{F}^* \;.
\end{equation}
For the specific Hermitian Yukawa ansatz we are considering,
 we find the following SM contributions:
\be
\varepsilon_K^{\mathrm{SM}} \simeq 1.8 \times 10^{-3}.
\ee
We see that the SM prediction for $\varepsilon_K$ is close to the measured 
value. However, a precise prediction cannot be made due to the hadronic 
and CKM uncertainties. 

Now let us turn to the supersymmetric contributions. The Higgs and 
the neutralino contributions are very suppressed and can be neglected. 
The chargino contribution to the $K^0-\bar{K}^0$ mixing is given 
by~\cite{chargino}
\begin{equation}
\label{susy:eK:m12char}
M_{12}^{\tilde{\chi}^\pm} = \frac{g^2}{768 \pi^2 m_{\tilde{q}^2}} \frac{1}{3}
M_{K}f_{K}^{2} B_1(\mu) \left(\sum_{a,b} K_{a2}^* 
(\delta^u_{LL})_{ab} K_{b1}\right)^2
\sum_{i,j} \vert V_{i1}\vert^2 \vert V_{j1} \vert^2 H(x_i,x_j),
\end{equation}
where $x_i=(m_{\tilde{\chi}_i^+}/m_{\tilde{q}})^2$, $K$ refers the CKM matrix, 
$a,b$ are the flavor indices, $i,j$ label the 
chargino mass eigenstates, and $V$ is the matrix that is used for
diagonalizing the chargino mass matrix.
The loop function $H(x_i, x_j)$ is given in Ref.\cite{chargino}.
However, the chargino contribution can be significant only if there is a 
large LL mixing in the up- sector, namely  Im$(\delta^u_{LL})_{21} 
\sim 10^{-3}$ and Re$(\delta^u_{LL})_{21} \sim 10^{-2}$~\cite{chargino}.
Such mixing can not be accommodated with the universal scalar masses assumption
(\ie, $\delta_i=1$). In this case, the values of the 
$\mathrm{Im}(\delta^u_{LL})$ are of order $10^{-6}$ which leads to a 
negligible chargino contribution to $\varepsilon_K$.
With non--universal soft scalar masses ($\delta_i \neq 1$) a possible 
enhancement in the chargino contribution is expected however, as we will 
show, this non--universality also leads to a larger enhancement in the 
gluino contribution.
So, the dominant SUSY contribution to the $K-\bar K$ mixing in 
this class of models will be provided by the gluino exchange diagrams.

The gluino contribution to the $\Delta S=2$ effective Hamiltonian is given 
by~\cite{ciuchini}
\begin{eqnarray}
&&H^{\Delta S=2}_{\rm eff}= \frac{-\alpha_s^2}{216
m^2_{\tilde{q}}} \Big\{(\delta^d_{12})^2_{LL} \Big(24 Q_1 x f_6(x)+ 
66 Q_1 \tilde{f}_6(x) \Big) + (\delta^d_{12})^2_{RR} \Big(24
\tilde{Q}_1 x f_6(x) \nonumber\\
&&+ 66 \tilde{Q}_1 \tilde{f}_6(x)\Big) 
+(\delta^d_{12})_{LL} (\delta^d_{12})_{RR}  \Big(504 Q_4 x
f_6(x) -72 Q_4 \tilde{f}_6(x) + 24 Q_5 x f_6(x)\nonumber\\
&&+ 120 Q_5 \tilde{f}_6(x)\Big)
+ (\delta^d_{12})^2_{RL} \Big(204 Q_2 x f_6(x) - 36 Q_3 x
f_6(x) \Big) + (\delta^d_{12})^2_{LR} \Big(204 \tilde{Q}_2 x
f_6(x)\nonumber\\
&& - 36 \tilde{Q}_3 x f_6(x) \Big)+(\delta^d_{12})_{LR} 
(\delta^d_{12})_{RL} \left(-132 Q_4
\tilde{f}_6(x) - 180 Q_5 \tilde{f}_6(x) \right)\Big\},
\label{m12k}
\end{eqnarray}
where $x=m_{\tilde{g}}^2/m^2_{\tilde{q}}$, $m_{\tilde{q}}$ is the average
squark mass, $m_{\tilde{g}}$ is the gluino mass, and the functions $f_6(x)$,
$\tilde{f}_6(x)$ are given by
\begin{eqnarray}
f_6(x) &=& \frac{6(1+3 x) \ln x + x^3 - 9 x^2 - 9 x + 17}{6(x-1)^5},\\
\tilde{f}_6(x) &=& \frac{6(1+3 x) \ln x + x^3 - 9 x^2 - 9 x + 17}{6(x-1)^5}.
\end{eqnarray}
The matrix elements of the operators $Q_i$ between the  $K$-meson states
in the vacuum insertion approximation (VIA), where $B=1$, can be found in 
Ref.\cite{Gabbiani}. The VIA 
generally gives only a rough estimate, so other methods, e.g. lattice QCD, 
are required to obtain a more realistic value. The matrix elements
of the renormalized operators can be written as~\cite{ciuchini}
\begin{eqnarray}
\langle \bar{K}^0 \vert Q_1(\mu) \vert K^0 \rangle &=& \frac{1}{3} M_K
f_K^2 B_1(\mu),\\
\langle \bar{K}^0 \vert Q_2(\mu) \vert K^0 \rangle &=& -\frac{5}{24}
\left(\frac{M_K}{m_s(\mu) + m_d (\mu)}\right)^2 M_K
f_K^2 B_2(\mu),\\
\langle \bar{K}^0 \vert Q_3 (\mu) \vert K^0 \rangle &=& -\frac{1}{24}
\left(\frac{M_K}{m_s(\mu) + m_d (\mu)}\right)^2 M_K
f_K^2 B_3(\mu),\\
\langle \bar{K}^0 \vert Q_4(\mu) \vert K^0 \rangle &=& -\frac{5}{24}
\left(\frac{M_K}{m_s(\mu) + m_d (\mu)}\right)^2 M_K
f_K^2 B_4(\mu)\;,
\end{eqnarray}
where $Q_i(\mu)$ are the operators renormalized at the scale $\mu$.
The expressions of the matrix elements of the operators $Q_{1-3}$ 
are valid for the operators $\tilde{Q}_{1-3}$ \cite{ciuchini}, and for 
$\mu =2$ GeV we have \cite{ciuchini}
\begin{eqnarray}
B_1(\mu) = 0.60, ~ B_2(\mu) = 0.66,~ B_3(\mu) = 1.05,~ B_4(\mu) = 1.03,~
B_5(\mu) = 0.73.
\end{eqnarray}
Using these values, the gluino contribution to the $K^0-\bar K^0$ can be 
calculated
via Eq.(\ref{deltamk}). As mentioned above,
for universal soft scalar masses the LL and RR insertions are generated 
only through the RG running and can be neglected. 
The LR and RL mass insertions appear at the tree level and may have 
tangible effects. It is worth mentioning that, the RL and LR mass insertions
contribute with the same sign in Eq.(\ref{m12k}) and for Hermitian A-terms
they are almost equal, so no cancellation between these two contributions 
occurs. 

\begin{figure}[t]
\begin{center}
\hspace*{-7mm}
\epsfig{file=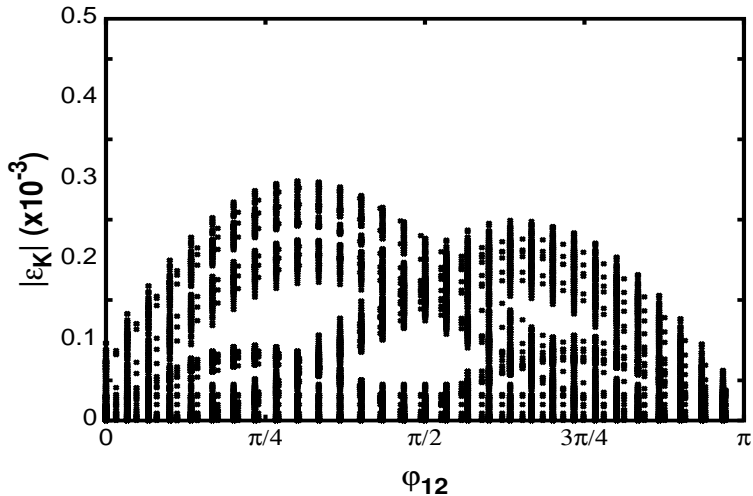,width=10cm,height=6.25cm}\\
\caption{{\small The gluino contribution to $\vert \varepsilon_K \vert$
as a function of $\varphi_{12}$ for 
$\delta_i=1$,  $\tan\beta=5$, and $m_0=m_{1/2}=200$ GeV.}}
\label{fig2}
\end{center}
\end{figure}

In Fig. 2 we plot the values of $\vert \varepsilon_K \vert$ versus 
the phase $\varphi_{12}$ for $\delta_i=1$ and the other parameters are chosen 
as in Fig. 1. From this figure, we conclude 
that the SUSY contribution with Hermitian Yukawa and universal soft scalar
masses can not account for the experimentally observed indirect CP
violation in the Kaon system. In Ref.\cite{ABKL} values for
$\varepsilon_K\sim 10^{-3}$ have been obtained but these values require 
light gaugino mass ($m_{1/2}\sim 100$ GeV) which is now excluded by the new
experimental limits on the mass of the lightets Higgs. Also it requires
that the magnitude of the off--diagonal entries of the $A$--terms should be
much larger (at least five times larger) than the diagonal ones, which
looks unnatural. 

A possible way to enhance $\varepsilon_K$ is to have non--universal
soft squark masses at GUT scale. As mentioned above, the soft scalar masses
are not necessarily universal in generic SUSY models and their 
non--universality is not constrained by the EDMs. Thus for 
$\delta_i \neq 1$ the mass insertion $(\delta^d_{12})_{RR}$ is enhanced and 
we can easily saturate $\varepsilon_K$ through the gluino contribution.
To see this more explicitly, let us consider the LL and RR squark mass matrices
in the super--CKM basis
\bea
\left(M^2_{\tilde{d}}\right)_{LL}&\sim &V^{d\dagger}~ M^2_{Q}~
V^{d },\nonumber\\
\left(M^2_{\tilde{d}}\right)_{RR}&\sim&V^{d\dagger}~(M^2_{D})^T~ 
V^{d }\;.
\eea
Due to the universality assumption of $M_Q^2$ at GUT scale, the matrix
$\left(M^2_{\tilde{d}}\right)_{LL}$ remains approximately universal and 
the mass insertions $(\delta_{12}^d)_{LL}$ are sufficiently small 
($\mathrm{Im}(\delta_{12}^d)_{LL} \sim 10^{-5}$). However, since the masses
of the squark singlets $M^2_D$ are not universal, Eq.(\ref{soft2}),
sizeable off--diagonal elements in $\left(M^2_{\tilde{d}}\right)_{RR}$
are obtained. We find that $\mathrm{Re}(\delta_{12}^d)_{RR}$ is enhanced
from $\sim 10^{-7}$ in the universal case ($\delta_i=1$) to 
$\sim 10^{-3}$ for $\delta_i\sim 0.7$ while the 
imagenary part remains the same, of order $10^{-7}$. Thus, in this case, 
we have 
$$\sqrt{\vert \mathrm{Im}\left((\delta^d_{12})_{LL} (\delta^d_{12})_{RR}
\right)\vert} \simeq \sqrt{\vert \mathrm{Re} (\delta^d_{12})_{RR}
\mathrm{Im}(\delta^d_{12})_{LL}\vert}\simeq 10^{-4}$$ 
which is the required value in order to saturate the observed result of 
$\varepsilon_K$~\cite{Gabbiani}.  

\begin{figure}[t]
\begin{center}
\hspace*{-7mm}
\epsfig{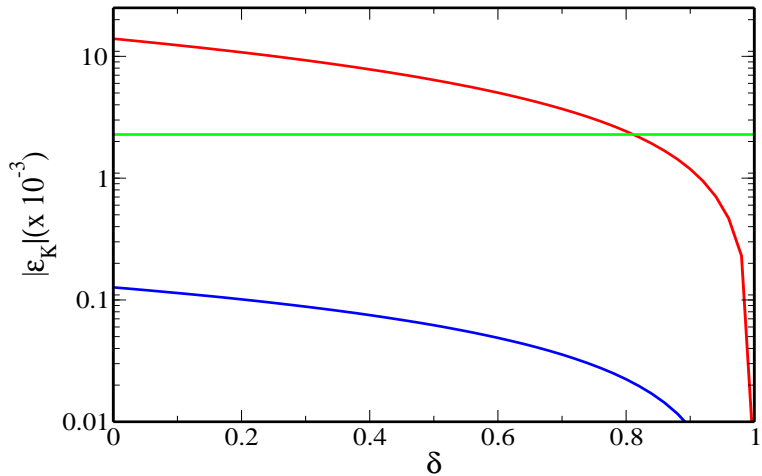}\\
\caption{{\small The value of $\vert \varepsilon_K \vert$
as a function of the parameters $\delta_1$ (upper curve) and
$\delta_2$ (lower curve) for $\tan\beta=5$, and $m_0=m_{1/2}=200$ GeV.}}
\label{fig3}
\end{center}
\end{figure}

In Fig. 3 we show the dependence of $\vert \varepsilon_K \vert$
on the parameters $\delta_i$. There, the two curves, from top to bottom, 
correspond to the values of $\vert \varepsilon_K \vert$ versus 
$\delta_1$ (for $\delta_2=1$) and $\delta_2$ (for $\delta_1=1$) respectively. 
As explained above, in this scenario the main contribution of 
$\varepsilon_K$ is due to 
LL and RR sectors and the LR sector has essentially no effect on  
$\varepsilon_K$. We also see that any non--universality between
the soft scalar masses of the third and the first two generations
can not lead to significant contribution to $\varepsilon_K$ and 
some spliting between the scalar masses of the first two generations
is necessary. This stems from the fact that the effect of the third 
generation on the mass insertion $(\delta_{12}^d)_{RR}$ is suppressed 
by $V_{13} \sim \mathcal{O}(10^{-2})$ while $V_{12}\sim \sin\theta_C$. 
Finally, as we can see from this figure, in order to avoid over saturation 
of the experimental value of $\varepsilon_K$, the parameter $\delta_1$ should 
be of order 0.8.

%
\subsection{\bf \normalsize Direct CP violation}

Next let us consider  SUSY contributions to $\varepsilon^{\prime}/\varepsilon$.
The ratio $\varepsilon^{\prime}/\varepsilon$ is a measure of 
direct CP violation in the $K \to \pi \pi$ decays and is given by
\begin{equation}
\varepsilon'/\varepsilon=-{\omega \over \sqrt{2}~\vert \varepsilon 
\vert~ {\rm Re}A_0}~ \left( {\rm Im} A_0 -{1\over \omega}~{\rm Im}A_2 \right) ,
\label{eprimeformula}
\end{equation}
where $A_{0,2}$ are the amplitudes for the $\Delta I=1/2,3/2$ transitions, 
and $\omega\equiv {\rm Re} A_2 /{\rm Re} A_0 \simeq 1/22 $ reflects
the strong enhancement of $\Delta I= 1/2$ transitions over those with
$\Delta I= 3/2$. Experimentally it has been found to be 
${\rm Re}(\varepsilon'/\varepsilon) \simeq 1.9 \times 10^{-3}$ 
which provides firm evidence for the existence of direct CP violation. 
The $\mathrm{Im} A_{0,2}$ are calculated from the general low energy
effective Hamiltonian for $\Delta S=1$ transition,
\begin{eqnarray}
&& H_{\rm eff}^{\Delta S=1}= \sum_i C_i(\mu)  O_i +h.c. \;,
\end{eqnarray}
where $C_i$ are the Wilson coefficients and the list of the relevant 
operators $O_i$ for this transition is given in 
Ref.\cite{buras2,emidio,buras3}. Let us recall here that these 
operators can be classified into three categories. The first category 
includes dimension six operators: $O_{1,2}$ which refer to 
the current-current operators, $O_{3-6}$ for QCD penguin operators and 
$O_{7-10}$ for electroweak penguin operators~\cite{buras2}. 
The second category includes dimension five operators: magnetic- 
and electric-dipole penguin operators $O_{g}$ and $O_{\gamma}$ which are 
induced by the gluino exchange~\cite{emidio}. 
The third category includes the only dimension 
four operator $O_Z$ generated by the $\bar{s}dZ$ vertex which is mediated 
by chargino exchanges~\cite{buras3}. In addition, one should take 
into account $\tilde{O}_i$ operators which are obtained from $O_i$ 
by the exchange $L \leftrightarrow R$.

In spite of the presence of this large number of operators that
in principle can contribute to $\varepsilon'/\varepsilon$, it is 
remarkable that few of them can give significant contributions.
As we will discuss below, this is due to the suppression of the matrix
elements and/or the associated Wilson coefficients of most of the operators.
The SM contribution to $\varepsilon'/\varepsilon$ is dominated by the 
operators $Q_6$ and $Q_8$, and can be expressed
as
\begin{equation}
\mathrm{Re}\left(\frac{\varepsilon^\prime}{\varepsilon}\right)^{\mathrm{SM}} = 
\frac{\mathrm{Im}\left(\lambda_t \lambda_u^*\right)}{\vert \lambda_u \vert}~ 
F_{\varepsilon^\prime},
\end{equation}
where $\lambda_i= V_{is}^* V_{id}$ and the function 
$F_{\varepsilon^\prime}$ is given in Ref.\cite{buras1}.
By using our Hermitian Yukawa in Eq.(\ref{yukawa}) we
get 
\be 
\varepsilon'/\varepsilon \simeq 7.5 \times 10^{-4}.
\ee
Again, the SM prediction is below the observed value. Nevertheless,
this estimat can not be considered as a firm conclusion for 
a new physics beyond the SM since there are significant hadronic 
uncertainties are involved.

The supersymmetric contribution to $\varepsilon'/\varepsilon$ depends
on the flavor structure of the SUSY model. It is known that, in a minimal 
flavor SUSY model, it is not possible to generate a sizeable contribution
to $\varepsilon'/\varepsilon$ even if the SUSY phases are assumed to be large. 
Recently, it has been pointed out that with non--degenerate $A$--terms 
the gluino contribution to $\varepsilon'/\varepsilon$ can naturally be 
enhanced to saturate the observed value~\cite{masiero,nonuniv}. 
Indeed, in this scenario, 
the LR mass insertions can have large imaginary parts and the chromomagnetic 
operator $O_g$ gives the dominant contribution to $\varepsilon'/\varepsilon$
\begin{equation}
\mathrm{Re}\left(\frac{\varepsilon^\prime}{\varepsilon}
\right)^g \simeq 
\frac{11 \sqrt{3}}{64 \pi^2 \vert \varepsilon \vert 
\mathrm{Re} A_0}~  \frac{m_s}{m_s+m_d} \frac{F_k^2}{F_{\pi}^3}~ m_K^2~
m_{\pi}^2~ \mathrm{Im}\left[ C_g - \tilde{C}_g\right] \;,
\end{equation}
where $C_g$ is the Wilson coefficient associated with the operator $O_g$,
given by
\be
C_{g} = \frac{\alpha_s \pi}{m_{\tilde{q}}^2} \left[
(\delta^d_{12})_{LL} 
\left(-\frac{1}{3} M_3(x) - 3 M_4(x) \right) +
(\delta^d_{12})_{LR} \frac{m_{\tilde{g}}}{m_s} 
\left(-\frac{1}{3} M_1(x) - 3 M_2(x) \right)
\right] \;,
\ee
where the functions $M_i(x)$ can be found in Ref.\cite{Gabbiani} and  
$x=m_{\tilde{g}}^2/m_{\tilde q}^2$.

Using these relations, one finds that in order to saturate 
$\varepsilon'/\varepsilon$ from the gluino contribution
the imaginary parts of the LR mass insertions for $x\simeq 1$ should satisfy
$\mathrm{Im}(\delta^d_{12})_{LR} \sim 10^{-5}$. Such values can easily
be obtained in this class of models. However, as mentioned above, in the 
case of Hermitian $A$--terms and Yukawa couplings we have 
$(\delta^d_{12})_{LR} \simeq (\delta^d_{12})_{RL}$, hence we get
$\mathrm{Im}\left[ C_g - \tilde{C}_g\right]\simeq \mathrm{Im}
\left[(\delta^d_{12})_{LR} - (\delta^d_{12})_{RL} \right] \simeq 10^{-6}$
which leads to a negligible gluino contribution to $\varepsilon'/\varepsilon$
\cite{ABKL}. It is worth noticing that, due to the universality assumption
of $M_{\tilde{Q}_L}^2$, the imaginary part of the mass insertion 
$(\delta^d_{12})_{LL}$ is of order $10^{-5}$. So its contribution to 
$C_g$ is negligible with respect to the LR one which is enhanced by the ratio
$m_{\tilde{g}}/m_s$. To achieve the required contribution to 
$\varepsilon'/\varepsilon$ from the LL sector, one has to relax this 
universality assumption to get $\mathrm{Im}(\delta_{12}^d)_{LL}\sim 10^{-2}$. 
However, as we will discuss below, in this case the chargino contribution 
is also enhanced and becomes dominant.

Now we turn to the chargino contributions. The dominant contribution 
is found to be due to the terms proportional to a single 
mass insertion~\cite{chargino}.
\begin{equation}\label{susy:char}
\mathrm{Re}\left(\frac{\varepsilon^\prime}{\varepsilon}\right)^{\chi^{\pm}}=
\mathrm{Im} \left(\sum_{a,b} K_{a2}^*
(\delta^u_{ab})_{LL} K_{b1}\right) \; F_{\varepsilon^\prime}(x_{q\chi})\;.
\end{equation}
The function $F_{\varepsilon^\prime}(x_{q\chi})$,
where $x_{q\chi}=m_{\tilde{\chi}^\pm}^2/m_{\tilde q}^2$,
is given in~\cite{chargino}.
We find that the contributions involving a double mass insertion, 
like those arising from the supersymmetric effective $\bar{s}dZ$, can not give
any significant contribution, however we take them into account.
The above contribution is dominated by 
$(\delta^u_{12})_{LL}$ and in order to account for $\varepsilon'/\varepsilon$
entirely from the chargino exchange the up sector has to employ
a large LL mixing. Again, with universal $M_{\tilde{Q}}^2$,  
$\mathrm{Im}(\delta^u_{12})_{LL}\sim 10^{-6}$ and the chargino 
contributions (as the gluino one) to $\varepsilon'/\varepsilon$ is 
negligible.  

Finally, we consider another possibility proposed by Kagan and Neubert 
to obtain a large contribution to $\varepsilon'/\varepsilon$ \cite{neubert}.
It is important to note that in the previous mechanisms to generate 
$\varepsilon'/\varepsilon$ one is tacitly assuming that $\Delta I=1/2$
transitions are dominant and that the $\Delta I=3/2$ ones are suppressed
as in the SM. However, in Ref.\cite{neubert} it was shown that it is
possible to generate a large $\varepsilon'/\varepsilon$ from the
$\Delta I=3/2$ penguin operators. This mechanism relies on the LL
mass insertion $(\delta_{21}^d)_{LL}$ and requires isospin violation
in the squark masses $(m_{\tilde{u}} \neq m_{\tilde{d}})$. 
In this case, the relevant $\Delta S=1$ gluino box diagrams lead to 
the effective Hamiltonian \cite{neubert}
\be
H_{\mathrm{eff}}= \frac{G_F}{\sqrt{2}} \sum_{i=3}^6 
\left( C_i(\mu) Q_i + \tilde{C}_i \tilde{Q}_i \right) + h.c. 
\ee
where 
\bea 
Q_1 &=& (\bar{d}_{\alpha} s_{\alpha})_{V-A}~
(\bar{q}_{\beta} q_{\beta})_{V+A},~~~~ 
Q_2 = (\bar{d}_{\alpha} s_{\beta})_{V-A}~
(\bar{q}_{\beta} s_{\alpha})_{V+A},\\
Q_3 &=& (\bar{d}_{\alpha} s_{\alpha})_{V-A}~
(\bar{q}_{\beta} q_{\beta})_{V-A},~~~~ 
Q_4 = (\bar{d}_{\alpha} s_{\beta})_{V-A}~
(\bar{q}_{\beta} q_{\alpha})_{V-A},
\eea
and the operators $\tilde{Q}_i$ are obtained from $Q_i$ by
exchanging $L \leftrightarrow R$. It turns out that the SUSY 
$\Delta I=3/2$ contribution to $\varepsilon'/\varepsilon$
is given by~\cite{neubert}
\be
\mathrm{Re}\left(\frac{\varepsilon^\prime}{\varepsilon}\right)^{\Delta I=3/2}
\simeq 19.2 \left(\frac{500 \mathrm{GeV}}{m_{\tilde{g}}}\right)^2 
B_8^{(2)} K(x_d^L, x_u^R, x_d^R)~ \mathrm{Im}(\delta^d_{12})_{LL} .
\ee
Here, $x_u^{L,R}= \left(\frac{m_{\tilde{u}_{L,R}}}{m_{\tilde{g}}}\right)^2$,
$x_d^{L,R}= \left(\frac{m_{\tilde{d}_{L,R}}}{m_{\tilde{g}}}\right)^2$,
$B_8^{(2)}(m_c)\simeq 1$ and the function $K(x,y,z)$ is given by
\be
K(x,y,z) = \frac{32}{27} \left[ f(x,y) - f(x,z) \right] 
+ \frac{2}{27}\left[ g(x,y) - g(x,z) \right],
\ee
where $f(x,y)$ and $g(x,y)$ are given in Ref.\cite{neubert}. 
It is clear from the above equation that for $m_{\tilde{d}_R}=
m_{\tilde{u}_R}$, the function $K(x_d^L, x_u^R, x_d^R)$ vanishes identically.
Thus, a mass splitting between the right--handed squark mass is 
necessary to get large contributions to $\varepsilon'/\varepsilon$ through 
this mechanism. Furthermore, the $\mathrm{Im}(\delta^d_{12})_{LL}$ has to 
be of order $\mathcal{O}(10^{-3} - 10^{-2})$ to saturate the observed 
value of $\varepsilon'/\varepsilon$. It is clear that, with universal 
$M^2_{\tilde{Q}}$, this contribution can not give any significant value
for $\varepsilon'/\varepsilon$. 

Now we relax the universality assumption of $M^2_{\tilde{Q}}$ at GUT scale
to enhance the mass insertion  $(\delta^d_{12})_{LL}$ and saturate the 
experimental value of $\varepsilon'/\varepsilon$. As mentioned above, 
the non--universality between the first two generation of $M^2_{\tilde{Q}}$
is very constrained by $\Delta M_K$ and $\varepsilon_K$. Therefore we just 
assume that the mass of third generation is given by $\delta_3 m_0$ while
the masses of the first two generations remain universal and equal to
$m_0$. This deviation from universality provides enhancement to both
$\varepsilon_K$ and $\varepsilon'/\varepsilon$. We have chosen the parameters
$\delta_i$ so that the total contributions of $\varepsilon_K$ from chargino 
and gluino are  consistent with the experimental limits.

\begin{figure}[t]
\begin{center}
\hspace*{-7mm}
\epsfig{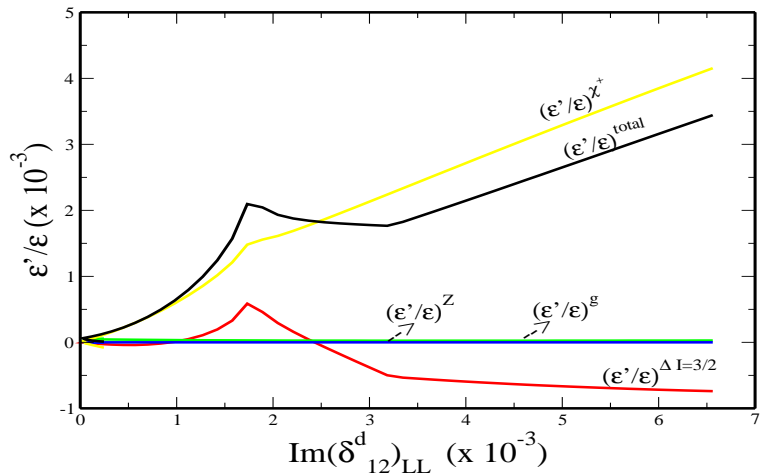}\\
\caption{{\small The  $\varepsilon'/\varepsilon$ contributions 
versus the Imaginary part of the mass insertion $(\delta^d_{12})_{LL}$.}}
\label{fig4}
\end{center}
\end{figure}

In Fig. 4 we present the different gluino and chargino contributions 
to the $\varepsilon'/\varepsilon$ and also the total contribution
versus the imaginary part of the mass insertion $(\delta^d_{12})_{LL}$.  
As explained above, there are two sources for the gluino contributions
to $\varepsilon'/\varepsilon$: the usual $\Delta I=1/2$ chromomagnetic
dipole operator and the new $\Delta I=3/2$ penguin operators.  
Additionally, there are two sources for the chargino contribution to 
$\varepsilon'/\varepsilon$: the usual gluon and electroweak penguin
diagrams with single mass insertion and the contribution from the SUSY
effective $\bar{s}dZ$ vertex. As can be seen from this figure, the dominant
contribution to  $\varepsilon'/\varepsilon$ is due to the chargino 
exchange with one mass insertion. It turns out that the chargino 
contribution with two mass insertions is negligible. 
As expected the gluino contribution via the chromomagnetic 
operator is also negligible due to the severe cancellation between the LR and RL
contributions. The contribution from the $\Delta I=3/2$ operators
does not lead to significant results for $\varepsilon'/\varepsilon$. It
even becomes negative (opposite to the other contributions) for 
$\mathrm{Im}(\delta^d_{12})_{LL} > 2.5 \times 10^{-3}$. 

From this figure we conclude that a non--universality among the soft
scalar masses is necessary to get large values of 
$\varepsilon'/\varepsilon$ and $\varepsilon_K$. 

%
\section{\bf \large CP violation in the $B$--system}
Recent results from the $B$--factories have confirmed, for the first time,  
the existence of CP violation in the $B$--meason decays. In particular, 
the measurements of the CP asymmetry in the $B_d \to \psi K_s$ decay~
\cite{2beta:babar,2beta:belle} have verified that CP is 
significantly violated in the 
$B$--sector. The time dependent CP--asymmetry $a_{\psi K_S}(t)$ is given by
\begin{equation}
a_{\psi K_S}(t) = \frac{\Gamma(B^0_d(t) \to \psi K_S) -
\Gamma(\bar{B}^0_d(t) \to \psi K_S)}{\Gamma(B^0_d(t) \to \psi K_S) +
\Gamma(\bar{B}^0_d(t) \to \psi K_S)}=- a_{\psi K_S} \sin(\Delta
m_{B_d} t)\;,
\end{equation}
where $\Delta M_{B_d}$ is the mass difference between the two mass 
eigenstates of the $B_d^0-\bar{B}_d^0$ system, given by
$\Delta M_{B_d} = 0.484 \pm 0.010~\mathrm{(ps)}^{-1}$~\cite{buras1}.
The measurements of this asymmetry is given by
\bea
a_{\psi K_s}&=& 0.59 \pm 0.14 \pm 0.05 \quad \quad
\mathrm{(BaBar)} \;, \nonumber \\
a_{\psi K_s}&=& 0.99 \pm 0.14 \pm 0.06 \quad \quad
\mathrm{(Belle)} \;.
\label{2beta:bounds}
\eea
This large CP asymmetry implies that CP is not an approximate 
symmetry in nature and that the CKM mechanism is the dominant source of 
CP violation.

In the framework of the SM, the unitarity of the CKM matrix implies 
the following relation
\be 
V_{ud} V_{ub}^* +V_{cd} V_{cb}^* + V_{td} V_{tb}^* =0 ,
\ee  
which can be represented as a unitarity triangle in the complex plane
$(\bar{\rho}, \bar{\eta})$~\cite{buras4}, where 
\be 
\bar{\rho}= \frac{1}{2} (1 + R_b^2 - R_t^2)~,~~ \bar{\eta}= \sqrt{ R_b^2 - 
\bar{\rho}},
\ee
and
\be 
R_b = \frac{\vert V_{ud} V_{ub}^*\vert}{\vert V_{cd} V_{cb}^* \vert}~,~~
R_t = \frac{\vert V_{td} V_{tb}^*\vert}{\vert V_{cd} V_{cb}^* \vert}.
\ee 
The angle $\beta$ of this unitarity triangle is related to 
the complex phase of element $V_{td}$ and is defined as~\cite{buras4}
\be
\sin 2 \beta = \frac{2 \bar{\eta} (1- \bar{\rho})}{(1-\bar{\rho})^2 
+\bar{\eta}^2}\;.
\ee
The asymmetry $a_{\psi K_s}$ in the SM is given by
$a_{\psi K_s}^{\mathrm{SM}}=\sin 2 \beta $. 
Using the Hermitian Yukawa ansatz of Eq.~(\ref{yukawa}), we find
that the SM contribution to $\sin 2 \beta$ is given by
\be
\sin 2 \beta^{\mathrm{SM}} \simeq 0.59.
\label{smbeta}
\ee
This result is in good agreement with the most recent world average
$ a_{\psi K_s}= 0.79 \pm 0.10$. Hence, any new contribution to the
CP asymmetry $ a_{\psi K_s}$ should be very limited. 
In supersymmetric theories the $\Delta B= 2$ transition, 
the off--diagonal element of the $B_d$ mass matrix, can be written as
\begin{equation}
M_{12}(B_d) =
\frac{\langle B^0_d \vert H_{\mathrm{eff}}^{\Delta B=2} \vert \bar{B}^0_d
\rangle}{2 m_{B_d}} = M_{12}^{\mathrm{SM}}(B_d) + M_{12}^{\mathrm{SUSY}}(B_d).
\end{equation}
The effect of supersymmetry can be described by a dimensionless parameter
$r_d^2$ and a phase $2 \theta_d$:
\begin{equation}
r_d^2 e^{2 i \theta_d} = \frac{M_{12}(B_d)}{M_{12}^{\mathrm{SM}}(B_d)},
\end{equation}
where $\Delta m_{B_d} = 2 \vert M_{12}^{\mathrm{SM}}(B_d)\vert r_d^2$.
Thus, in the presence of SUSY contributions, the
CP asymmetry $B_d \to \psi K_s$ is modified, and now we have 
\begin{equation}
a_{\psi K_S} = \sin (2 \beta + 2 \theta_d)\;.
\end{equation}
Therefore, the measurement
of $a_{\psi K_S}$ would not determine $\sin 2 \beta$ but rather $\sin(2 \beta
+ 2 \theta_d)$, where
\begin{equation}
2 \theta_d = \mathrm{arg}\left(1+\frac{M_{12}^{\mathrm{SUSY}}(B_d)}
{M_{12}^{\mathrm{SM}}(B_d)}\right).
\end{equation}
From Eqs. (\ref{2beta:bounds}) and (\ref{smbeta}) we find that the 
allowed range for $\sin 2 \theta_d$ is as follow
\be
-0.22 \lsim \sin 2 \theta_d \lsim 0.8.
\label{2thetarange}
\ee
In the following,  we will analyse the impact of the SUSY flavor 
off--diagonal phases on the allowed values of $a_{\psi K_S}$ and 
the implication of the above $\sin 2 \theta_d$ constraints for the SUSY model
we are considering. The effective Hamiltonian for $\Delta B=2$ 
processes can be expressed as
\begin{equation}
H^{\Delta B=2}_{\rm eff}=\sum_{i} C_i(\mu) Q_i + h.c. \;,
\end{equation}
where the Wilson coefficients and the relevant operators $Q_i$ can 
be found in Ref.\cite{Ali,masiero2}. The $M^{\mathrm{SUSY}}_{12}(B_d)$
receives significant contributions from the gluino and chargino 
exchange box diagrams (we also included the charged Higgs contribution 
in our numerical analysis).
The gluino contribution $M_{12}^{\tilde g}(B_d)$ can be obtained from  
Eq.(\ref{m12k}) by changing $\delta^d_{12} \to \delta^d_{13}$, 
$M_K \to M_{B_d}$, $f_K \to f_{B_d}$, $m_s \to m_b$. The 
chargino contribution is given by \cite{BBMR}
\begin{equation}
M_{12}^{\chi}(B_d) = \frac{\alpha_w^2}{24} f_{B_d}^2 B_{B_d}
\eta^{-6/23} M_{B_d} \sum_{h,k=1}^6 \sum_{i,j=1}^2 \frac{1}{m^2_{\tilde
{\chi}_j}} \mathcal{A}_{jki} \mathcal{A}_{ihj} 
G'(x_{\tilde{u}_k \tilde{\chi}_j}, x_{\tilde{u}_h \tilde{\chi}_j}, 
x_{\tilde{\chi}_i \tilde{\chi}_j}),
\end{equation}
where the function $\mathcal{A}_{ijk}$ and $G'(x,y,z)$ can be found
in Ref.\cite{BBMR}. In most of the parameter space, we found 
that the chargino contribution gives the dominant effect to 
the CP asymmetry $\sin 2 \theta_d$, while the gluino is sub-dominant.
This result can be understood by using the mass insertion method.
The gluino amplitude receives a leading contribution from 
$(\delta^d_{13})_{LL}$, since the mass insertions  
$(\delta^d_{13})_{LR}$ and $(\delta^d_{13})_{RR}$ are much smaller.
However, for $m_0=m_{1/2}\sim 200$ GeV, the $\vert (\delta^d_{13})_{LL}
\vert \sim 10^{-3}$ which is two orders of magnitude below the 
required value to saturate the experimental value of $a_{\psi K_S}$~
\cite{masiero2}, so that one has negligible gluino contribution to 
$\sin 2 \theta_d$. The chargino amplitude is proportional to 
$(\delta^u_{13})_{LL}$ which can be enhanced by a light stop mass. 

\begin{figure}[ht]
\begin{center}
\hspace*{-7mm}
\epsfig{file=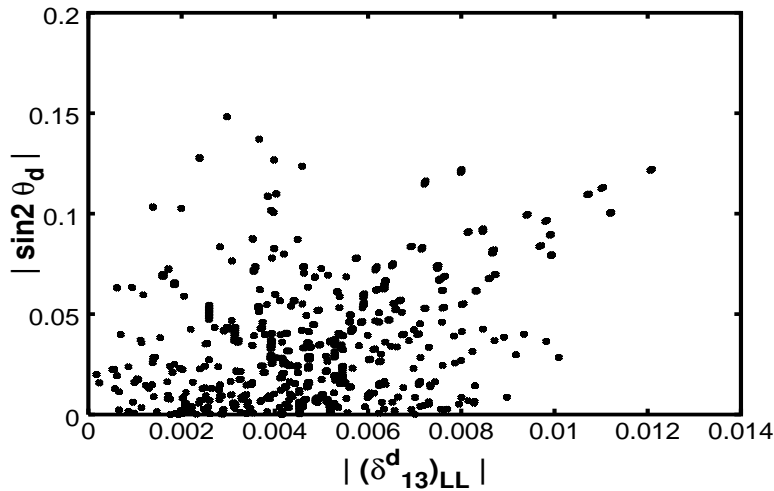,width=10cm,height=6.25cm}\\
\caption{{\small The SUSY contribution to $\sin 2 \theta_d$ and 
versus the $(\vert \delta^d_{13})_{LL} \vert$.}}
\label{fig5}
\end{center}
\end{figure}
Our results for the total SUSY contribution to the CP asymmetry 
$\sin 2 \theta_d$ as a function of the $\vert (\delta_{13}^d)_{LL} \vert$
are presented in Fig. 5. Here we have also assumed $m_0 =m_{1/2}=200$ GeV
and $\tan \beta = 5$. The parameters $A_{ij}$, $\varphi_{ij}$ and 
$\delta_{1,2}$ are varied in their allowed regions fixed by the experimental
limits on the EDMs, $\varepsilon_K$ and $\varepsilon/\varepsilon$. 
As expected, in order to have significant SUSY contributions to the 
CP asymmetry $a_{\psi K_S}$ a large mixing in the LL sector is required
(in order to enhance the dominant chargino contribution). 
However, such mixing is not allowed in our model with Hermitian Yukawas,
as Fig. 5 confirmed. This, in fact, is due to 
the $\varepsilon_K$ constraint that severely cuts off any enhancement
though the non--universality of the soft scalar masses.     

Also as can be seen from Fig. 5, the predicted values of $\sin 2 \theta_d$
reside within the allowed range defined in Eq.(\ref{2thetarange}). Hence, 
in this class of models, the SM gives the leading contribution
to the CP asymmetry $a_{\psi K_S}$, while the supersymmetric contribution 
is very small. Therefore, there is no constraint can be set on the
non--universality of this class of models by the recent BaBar and Belle 
measurements.

We conclude this section with some remarks on the CP asymmetry in  
$B \to X_s \gamma$. It is known that the SM prediction for the CP 
asymmetry is very small, less than
$1\%$. Thus, the observation of sizeable asymmetry in this decay
would be a clean signal of new physics. The most recent result
reported by the CLEO collaboration for the CP asymmetry in  these decays is
$ -27\% < A_{CP}^{b\to s \gamma}  < 10 \%$~\cite{cleo}.   
The SUSY predictions for $A_{CP}^{b\to s \gamma}$ are strongly
dependent on the flavor structure of the soft breaking terms.
In the universal case, as in the minimal supergravity models, 
the asymmetry is found to be less than $2\%$~\cite{Goto}. 
However, it was shown that the non--universal $A$--terms can result 
in a large CP--asymmetry in the $B \to X_s \gamma$~\cite{Bailin} and
these effects are correlated with the EDMs. 

The enhancement of $A_{CP}^{b\to s \gamma}$ is due to
important contributions from gluino--mediated diagrams, in this
scenario, in addition to the usual chargino and charged Higgs contributions.
The expression for the asymmetry  $A_{CP}^{b\to s \gamma}$, corrected to
next--to--leading order is given by \cite{neubert2}
\bea
A_{CP}^{b\to s \gamma} &=& \Frac{4\alpha_s(m_b)}{9 \vert C_7 \vert^2}
\biggl\{\biggl[\frac{10}{9} - 2 z~ (v(z)+b(z,\delta))\biggr] 
\mathrm{Im}\left[C_2 C_7^* + \tilde{C}_2 \tilde{C}^*_7\right]
+ \mathrm{Im}\left[C_7 C^*_8+\tilde{C}_2 \tilde{C}^*_8\right]\nonumber\\ &+& 
\frac{2}{3} z~b(z,\delta) \mathrm{Im}\left[C_2 C_8^8+ \tilde{C}_2 
\tilde{C}^*_8\right] \biggr \},
\label{asymmetry}
\eea
where $z=m_c^2/m_b^2$. The functions $v(z)$ and $b(z,\delta)$
and the Wilson coefficients $C_i$ can be found in Ref.\cite{neubert2,Bailin}.
In the EDM-free models we are considering, we found that
the flavor dependent phase $\varphi_{23}$ gives a large contribution
to the CP asymmetry (since the Wilson coefficients are proportional 
to $(\delta_{23}^d)_{LR}$ which receives a dominant contributions 
from $A_{23}$ entry~\cite{Bailin}). The effect of the other flavor 
dependent phases on $A_{CP}^{b \to s \gamma}$ is found to be very small.

In our SUSY model with  Hermitian $A$--terms, we found that the gluino 
contribution to CP asymmetry  $A_{CP}^{b\to s \gamma}$ can be as large as 
$10\%$, which can be accessible at the $B$ facrories.

\section{\bf \large Conclusions}
In this paper, we have analysed the CP violation in supersymmetric
models with Hermitian Yukawa and trilinear couplings. We emphasized that
in most of the parameter space of this class of models the EDM
of the neutron and mercury atom are two orders of magnitude smaller 
than the experimental limits. Furthermore, we have shown that Hermitian
Yukawas can naturally be implemented in SUSY models with $SU(3)$ flavor 
symmetry.

We have studied the CP violation in the Kaon system. We found that in order
to saturate $\varepsilon_K$ a small non--universality between the squark 
soft masses is required. We investigated the effects of the EDM--free, flavor 
off--diagonal, phases on the direct CP violation observable 
$\varepsilon'/\varepsilon$. A large SUSY contribution to this observable
is possible via the chargino contribution.

Finally, we considered the $a_{\psi K_S}$ and $A_{CP}^{b\to s \gamma}$ 
CP asymmetries in $B$--meason decays.
We verified that the SM 
contribution to $a_{\psi K_S}$ is in agreement with the recent measurements
by BaBar and Belle, while the SUSY contributions are very small and hence
no further constraint is imposed on the non--universality of these models.
In contrast, with $A_{CP}^{b\to s \gamma}$ the SM contribution is negligible
and the SUSY contribution can be as large as $\pm 10\%$ which can
be accessible at B--factories.

%
\section*{\bf \large Acknowledgements}

I would like to thank S. Abel, O. Lebedev and A. Teixeira for very useful 
discussion. This work was supported by the PPARC.


\end{document}